\documentclass{ieeeaccess}

\usepackage{cite}
\usepackage{amsmath,amssymb,amsfonts}
\usepackage{algorithmic}
\usepackage{graphicx}
\usepackage{textcomp}

\usepackage{epstopdf}
\usepackage{subfigure}
\usepackage{color}
\usepackage{longtable}
\usepackage{url}
\usepackage{multirow}
\usepackage{eurosym}

\def\BibTeX{{\rm B\kern-.05em{\sc i\kern-.025em b}\kern-.08em
    T\kern-.1667em\lower.7ex\hbox{E}\kern-.125emX}}

\begin{document}

\doi{10.1109/ACCESS.2024.3406392}

\title{Low Cost Carriers induce specific and identifiable delay propagation patterns: an analysis of the EU and US systems}

\author{
\uppercase{Sofia Gil-Rodrigo}\authorrefmark{1} and \uppercase{Massimiliano Zanin}\authorrefmark{1}
}

\address[1]{Instituto de F\'isica Interdisciplinar y Sistemas Complejos IFISC (CSIC-UIB), Campus UIB, 07122 Palma de Mallorca, Spain (e-mail: massimiliano.zanin@gmail.com)}

\tfootnote{This project has received funding from the European Research Council (ERC) under the European Union's Horizon 2020 research and innovation programme (grant agreement No 851255). This work was partially supported by the Mar\'ia de Maeztu project CEX2021-001164-M funded by the  MICIU/AEI/10.13039/501100011033. S.G.-R. acknowledges funding from CSIC, JAE program.
}

\markboth
{Gil-Rodrigo {\it et al.}: Low Cost Carriers induce specific and identifiable delay propagation patterns}
{Gil-Rodrigo {\it et al.}: Low Cost Carriers induce specific and identifiable delay propagation patterns}

\corresp{Corresponding author: Massimiliano Zanin (e-mail: massimiliano.zanin@gmail.com).}

\begin{abstract}
The impact of air transport delays and their propagation has long been studied, mainly from environmental and mobility viewpoints, using a wide range of data analysis tools and simulations. Less attention has nevertheless been devoted to how delays create meso-scale structures around each airport. In this work we tackle this issue by reconstructing functional networks of delay propagation centred at each airport, and studying their identifiability (i.e. how unique they are) using Deep Learning models. We find that such delay propagation neighbourhoods are highly unique when they correspond to airports with a high share of Low Cost Carriers operations; and demonstrate the robustness of these findings for the EU and US systems, and to different methodological choices. We further discuss some operational implications of this uniqueness.
\end{abstract}


\titlepgskip=-15pt
\maketitle

\section{Introduction}

Delays are one of the major topics of research in air transport, due to their profound implications in the cost-efficiency \cite{cook2009cost} and safety of the system \cite{duytschaever1993development}, and their negative impact on the environment \cite{carlier2007environmental}. To illustrate, the Federal Aviation Administration estimates that US flight delays cost \$$22$bn yearly: getting rid of delays would thus allow to pay one third of the whole national health care system of Spain (\euro$72.8$bn in $2017$). Additionally, $1$ minute of ground delay implies between $1$ to $4$ kg of fuel consumption, one order of magnitude higher in the case of airborne delays \cite{carlier2007environmental}. If some delays are the unavoidable result of external perturbations, as e.g. of adverse meteorological events, the same cannot be said of their propagation, also known as reactionary or secondary delays. These propagation instances are the result of the high level of optimisation of the system, and of the limited resources available to airlines, airports, and air traffic managers. Not surprising, the appearance of delays and their propagation has been studied using a plethora of complementary approaches, from the analysis of the local dynamics of individual flights and airports \cite{beatty1999preliminary, liu2008estimation, ahmadbeygi2008analysis}; the use of large-scale synthetic models \cite{ahmadbeygi2008analysis, fleurquin2013systemic, pyrgiotis2013modelling, baspinar2016data, zhang2020simulation}; to functional network representations inspired by statistical physics and neuroscience \cite{zanin2015can, zanin2017network, du2018delay, mazzarisi2020new, pastorino2021air}.

One interesting aspect that has recently been raised is how identifiable \cite{zanin2023identifiability} delays are, or, in other words, whether it is possible to identify an airport by only looking at the delays it experiences \cite{ivanoska2022assessing}. Whenever this is possible, it means that the evolution of delays throughout different days is similar - or at least that it is more similar than what observed in other airports. Most importantly, this also means that delays are predictable and avoidable - if, for instance, an airport experiences the same spike in the average delay at a given hour every day, resources may be deployed to minimise it, or flights may have to be rescheduled.

If Ref. \cite{ivanoska2022assessing} focused on the dynamics of delays at each airport, here we adopt a more meso-scale approach and evaluate the identifiability of the structure induced by the propagation of delays in its neighbourhood. In short, given a target airport, we identify the subset of airports to which it strongly connects, for then recovering how delays have propagated between them on a daily basis. The result is then a functional network per day, describing how delays have propagated in the neighbourhood of the target airport. Networks coming from pairs of airports are then classified using Deep Learning models, and specifically Graph Isomorphism Neural Networks (GIN) \cite{Xu_Hu_Leskovec_Jegelka_2019}. The accuracy of the classification then quantifies how unique, or identifiable, those networks are, and therefore how unique is the structure of the propagation of delays near those airports.

When the proposed approach is applied to large data sets of flights of Europe and US, results indicate that these delay propagation neighbourhoods are generally not identifiable, possibly due to the high variance in delays across different days. At the same time, some airports stand out for being highly unique, e.g. M\'alaga-Costa del Sol Airport, Alicante-Elche Miguel Hern\'andez Airport, and London Stansted Airport in Europe; and San Jos\'e International Airport, Fort Lauderdale-Hollywood International Airport, and Chicago O'Hare International Airport in the US. We demonstrate that such uniqueness is connected to the share of flights operated by Low Cost Carriers, and hence how these induce specific delay propagation patterns. We also show how these results are independent on several methodological choices, like the way functional networks are reconstructed and how the classification is performed.

The remainder of the text starts by introducing the main methods of the analysis, specifically: the considered data set (Sec. \ref{sec:dataset}), the delay network reconstruction process (Sec. \ref{sec:delaynet}), the Deep Learning model (Sec. \ref{sec:dl}), and the topological metrics used to characterise the networks (Sec. \ref{sec:top_metrics}). We present the main results for the EU system in Sec. \ref{sec:results}, discussing how these are generalisable with respect to the network reconstruction process (Sec. \ref{sec:gen_func}), and how they are also valid in the case of US (Sec. \ref{sec:gen_US}). We finally discuss the operational implications of this work and draw some conclusions in Sec. \ref{sec:concl}.

\section{Methods}
\label{sec:methods}

\subsection{Data sets}
\label{sec:dataset}

Data about European flights have been extracted from the EUROCONTROL's R\&D Data Archive, a public repository of historical flights made available for research purposes and freely accessible at \url{https://www.eurocontrol.int/dashboard/rnd-data-archive}. It includes information about all commercial flights operating in and over Europe, completed with flight plans, radar data, and associated airspace structure. From a temporal point of view, it includes data for four months (i.e. March, June, September and December) of five years (2015-2019). Note that data for year 2020 and following have been discarded, due to the anomalies introduced by the COVID-19 pandemics. In this study we focused on the $35$ largest airports in Europe according to the number of passengers. Tab. \ref{tab:stats_airp} in Appendix reports the full list along with the corresponding number of landing operations.

For each flight landing at these $35$ airports, its delay has been calculated as the difference between the actual (from the ATFM-updated flight plan) and the planned (according to the last filed flight plan) landing times. Afterwards, for each airport, flights have been grouped according to the actual landing hour, and the average delay per hour has been calculated. Subsequently, these time series have been split in windows of 24 hours, in order to obtain the evolution of the average hourly delay per airport per day. The result is thus a total of $744$ time series of length $24$ per airport.

Similar information has further been obtained for the $35$ largest US airports from the Reporting Carrier On-Time Performance database of the Bureau of Transportation Statistics, U.S. Department of Transportation, freely accessible at \url{https://www.transtats.bts.gov}. This database contains information about flights operating in US airports, including departure and arrival time (both scheduled and executed), and consequently the associated delays. Data here considered cover years 2015 to 2019 (both included), thus yielding a larger set of $1,825$ time series per airport. A list of considered airports is reported in Tab. \ref{tab:stats_airp_US} in Appendix.

\subsection{Delay propagation network reconstruction}
\label{sec:delaynet}

The previously extracted time series are highly non-stationary, as delays appear with clear temporal trends - e.g. seldom in the morning but frequently at midday and afternoon. As stationarity is a necessary requirement for detecting instances of delay propagation, we firstly detrended them using a Z-Score approach. Specifically, given a time series $x(t)$ (with $t \in [0, \ldots, 23]$), its detrended version $x'(t)$ is given by:

\begin{equation}
    x'(t) = \frac{ x(t) - \mu(t) }{\sigma(t)},
\end{equation}

where $\mu(t)$ is the average delay observed at the same hour and in the same day of the week, and $\sigma(t)$ the corresponding standard deviation. In other words, the result is a value representing how the delay at a given time deviates from what expected in similar days and hour.

The detection of the propagation of delays between a pair of airports is usually performed through a functional metric, i.e. a metric assessing the presence of a synchronisation or of an information transfer in the corresponding time series. While many alternatives have been explored in the literature, including for instance Granger Causality \cite{granger1969investigating} or Transfer Entropy \cite{schreiber2000measuring, paluvs2001synchronization}, we are here constrained by the limited length of the time series (i.e. 24 points); complex and non-linear metrics, which usually require long time series to yield reliable results, may not be suitable. Consequently, we here initially resorted to a simple Pearson's linear correlation, which has already successfully been used in detecting delay propagation \cite{wang2020comparative, chen2020community, wang2022timescales}; the use of other metrics will further be explored below.

Given a target airport $a$, we firstly extract the set of $n_{airps} - 1$ airports with which it had the largest number of flights - or, in other words, the subset of airports with which it is most strongly connected. We then reconstruct the functional network of delay propagation between these $n_{airps}$ airports (i.e. the target airport $a$ plus its $n_{airps} - 1$ neighbours), which is fully described by a weighted adjacency matrix $\mathcal{W}$ of size $n_{airps} \times n_{airps}$, where the element $w_{i,j}$ is given by the absolute value of the linear correlation between the detrended time series of airports $i$ and $j$. Finally, links whose weight $w$ is below a threshold $\tau$ are deleted, to avoid including correlations with no statistical significance - the optimal value of $\tau$ will be obtained in Sec. \ref{sec:hp}. In synthesis, given a target airport and a day, the result of this processing is a weighted network describing how synchronised is the delay evolution, during that day, among its neighbours. A total of $35 \times 744 = 26,040$ networks have been extracted for EU; $35 \times 1,825 = 63,875$ networks in the case of US.

\subsection{Deep Learning classification}
\label{sec:dl}

Once the delay propagation networks are obtained, we aim to find whether they have similar structures across different airports, or, on the other hand, whether they present consistent differences that allow for their identification. Such assessment is here performed through a classification task for each pair of airports, in which a model is trained to correctly label which airport a given network corresponds to; the higher the obtained classification score, the higher is the identifiability of the delay propagation networks under study.

The classification is performed using a Graph Isomorphism Neural Network (GIN) \cite{Xu_Hu_Leskovec_Jegelka_2019}. The origin of this type of neural network can be traced back to the Convolutional Neural Networks (CNNs) \cite{Wu_Pan_Chen_Long_Zhang_Yu_2021}, i.e. models capable of identifying local features and learning patterns through a structure that includes convolutional layers, pooling layers, and fully connected layers; and through a back-propagation training algorithm~\cite{Wu_Pan_Chen_Long_Zhang_Yu_2021, Yamashita_Nishio_Do_Togashi_2018}. For this reason, CNNs are a well-established model for object recognition and classification tasks when the input data have a grid shape, such as in the case of images~\cite{Yamashita_Nishio_Do_Togashi_2018}. Some situations nevertheless require representing the data as graphs, as is e.g. the case for the interactions between the elements of a system. Graph Neural Networks (GNNs), and Convolutional Graph Neural Networks (ConvGNNs) in particular, were developed to address this type of problems, generalising the convolution operation from a grid to a graph~\cite{Wu_Pan_Chen_Long_Zhang_Yu_2021}. 

At their core, GNNs operate through recursive neighbourhood aggregations, so that after $k$ iterations each node has a new representation that captures information about its $k$ nearest neighbours; and pooling operations, to synthesise a representation of the entire network. There are many different variants of GNNs and, as demonstrated in Ref.~\cite{Xu_Hu_Leskovec_Jegelka_2019}, GNNs can be as powerful as the Weisfeiler-Lehman (WL) test in terms of graph classification. Such test aims to determine if two graphs are topologically equivalent, meaning that they have the same connectivity and differ only by a permutation of their nodes, and it is a necessary but not sufficient condition for isomorphism~\cite{Sato_2020}. In Ref.~\cite{Xu_Hu_Leskovec_Jegelka_2019} an architecture called Graph Isomorphism Neural Network (GIN) is developed to reach this limit.
GINs are therefore the appropriate tool for classifying the delay propagation networks since they are able to capture the concept of isomorphism and have been proven to be a powerful tool for graph classification tasks \cite{Xu_Hu_Leskovec_Jegelka_2019}. We here leveraged the implementation from the Python library \textit{PyTorch} \cite{paszke2019pytorch}, modified to have the desired structure: three convolutional layers and two linear fully connected layers, with dimensions of respectively $(h,h)$ and $3h$, with $h=32$. 

The classification accuracy for a given airport pair is calculated starting from the corresponding neighbourhood propagation networks previously obtained (see Sec. \ref{sec:delaynet}), and by dividing the dataset into ten parts to implement a $k$-fold cross-validation technique. Consequently, we iterate the classification $k=10$ times, where, in each iteration, one different fold corresponds to the test set and the other nine to the train set. We perform the training and testing of the GIN model using a batch size of 64 networks, meaning that the GIN parameters are updated each time it processes 64 instances. Finally, for each pair of airports, this whole process is repeated $20$ times, to account for the stochastic nature of the training process; the final score is obtained as the average accuracy obtained throughout the $200$ training/evaluation rounds.

\subsection{Topological analysis of networks}
\label{sec:top_metrics}

In order to understand the characteristics of the extracted delay propagation networks, these have been analysed from the lens of complex network theory and through a set of topological metrics, i.e. metrics that describe specific aspects of their structure. These metrics will further be used to perform a classification task using a standard Machine Learning model, in order to evaluate the performance of DL ones. While many reviews on this topic have been published, see for instance \cite{boccaletti2006complex, costa2007characterization}, for the sake of completeness we here present some basic definitions.

\begin{itemize}
    \item {\it \# links}, or the number of links after applying a threshold $\tau$ - the role of this threshold will be discussed in Sec. \ref{sec:hp}.
    
    \item {\it Median weights}, i.e. median of the weight of links, here of the absolute value of the correlation coefficient between the delay time series.

    \item {\it Assortativity}, i.e. propensity of links to connect nodes of similar degrees \cite{newman2002assortative}, i.e. with similar number of connections. It is calculated as the Pearson's linear correlation between the degree of nodes at each end of each link.

    \item {\it Transitivity}, or the frequency of triangles in the network, measured as the ratio between the number of closed triangles and of connected triplets of nodes \cite{serrano2006clustering}.

    \item {\it Efficiency}, i.e. the normalised sum of the inverse of the distances between every pair of nodes \cite{latora2001efficient}. This metric measures how easily information is transmitted in the network; and, in the case at hand, how easily delays can propagate.

    \item {\it Modularity}, a metric representing the tendency of a network to organise into communities, or groups of nodes strongly connected between them and loosely connected to the remainder of the network \cite{fortunato2010community}. Communities have here been obtained using the Louvain algorithm \cite{blondel2008fast}.

    \item {\it Small-worldness}, ratio between the clustering coefficient and the characteristic path lengths, normalised according to the values expected in random equivalent networks \cite{humphries2008network}.

    \item {\it Network Information Content} (NIC), metric assessing the presence of regular structures in the network; it is based on an iterative process of merging pairs of nodes, and on measuring the amount of information lost in the process \cite{zanin2014information}.
\end{itemize}

Note that metrics like the transitivity, efficiency and NIC depend on the number of nodes and of links in the network. To illustrate, the higher the number of links in a network, the more probable is to find triangles, independently of the underlying structure. In order to normalise these metrics and allow comparisons between networks of different densities, we resort to a null model composed of random Erd\H{o}s-R\'enyi equivalent networks of the same number of nodes and links. The original metric is then expressed through the corresponding Z-Score, defined as: $z_M = (m - \overline{r_M}) / \sigma_{r_M}$, with $\overline{r_M}$ and $\sigma_{r_M}$ respectively being the mean and standard deviation of the metric $M$ calculated on the random networks - see Refs. \cite{maslov2002specificity, zanin2018studying} for further discussions.

\section{Results}
\label{sec:results}

\subsection{Hyperparameters' tuning}
\label{sec:hp}

As a prior step to the analysis of the obtained results, it is necessary to discuss the role of a few hyperparameters that affect the classification process. The first one is the size of each sub-network, i.e. $n_{airps}$. This hyperparameter controls a balance. On one hand, if the network contains very few airports, little information is available for the GIN algorithm to learn the classification, and therefore the yielded score will be low. On the other hand, in the limit that all $35$ airports are included, all sub-networks become the same and equal to the full propagation network, again preventing any meaningful learning. A maximum is therefore to be expected for intermediate values of $n_{airps}$.
A similar dynamics is to be expected in the case of the threshold $\tau$. A very large value of this hyperparameter implies that few links, if any, are included in the final sub-networks. On the other hand, it may be expected that including all available information, i.e. for $\tau = 0$ or performing no pruning, should favour the classification process. This is nevertheless not always the case: as well known in machine learning, performing a feature selection, i.e. deleting those features that convey little or no information, can help the algorithms to converge faster to a better solution \cite{kira1992practical, guyon2003introduction}. In the case at hand, this is equivalent to deleting those propagation links that are associated to small correlations, and that may therefore only represent noise.
The third hyperparameter of interest is the number of epochs $n_{epochs}$ during which the GNN model is trained. Low values of $n_{epochs}$ may prevent the algorithm to reach a valid solution; on the other hand, very large values may eventually lead to an overfitting.

The evolution of the classification score, measured through the mean accuracy, is reported in Fig. \ref{fig:hyperp} as a function of these three hyperparameters. For the sake of clarity, we report three pairs of European airports that have manually been selected to represent a variety of scenarios. It can be appreciated that the best classifications are obtained for $n_{airps} \approx 15$ and $\tau \approx 0.35$, in agreement with the previous discussion. Regarding $n_{epochs}$, values smaller than $10^2$ clearly hinder the classification process, while the best results are obtained for values larger than $10^3$. Consequently, the following values will be used in all subsequent analyses: $n_{airps} = 15$, $\tau = 0.35$ and $n_{epochs} = 5 \cdot 10^3$.

\Figure[t!][width=0.9\textwidth]{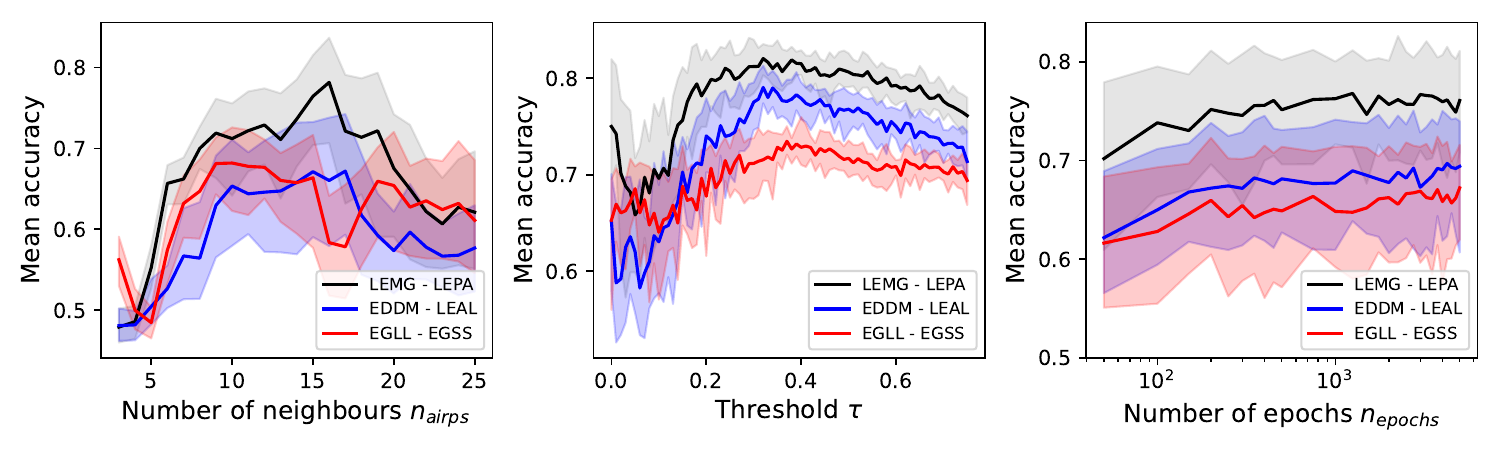}
{Tuning of the hyperparameters of the classification. The three panels report the evolution of the average accuracy of the classification as a function of: (left) the number of neighbours $n_{airps}$ composing the sub-networks; (centre) the threshold $\tau$; and (right) the number of epochs $n_{epochs}$. Solid lines depict the evolution of the average classification score for three pairs of airports (see legends), and transparent bands to the corresponding $10-90$ percentiles. Unless otherwise stated, $n_{airps} = 15$, $\tau = 0.35$, and $n_{epochs} = 5,000$. \label{fig:hyperp}}

\subsection{Identifiability of delay propagation networks}
\label{sec:res_class}

Once all hyperparameters have been set, it is now possible to analyse the classification results, and specifically the accuracy obtained when comparing the delay propagation neighbourhoods of each pair of airports. While all scores for the European airports are presented in the left panel of Fig. \ref{fig:all_pairs} in Appendix, for the sake of clarity the box plot of Fig. \ref{fig:boxplot} depicts the distribution corresponding to each airport; in other words, it presents an overview of how identifiable are the delay propagation patterns around a given airport. Highly heterogeneous behaviours can be observed: while some airports, as e.g. M\'alaga-Costa del Sol (LEMG) have delay propagation neighbourhoods that are easy to identify, many yield accuracy scores close to $0.5$, i.e. compatible with a random (or uninformed) classification. In order to simplify the interpretation of the results, Tab. \ref{tab:top_airps} reports a list of the top-6 European airports in terms of average accuracy; and Tab. \ref{tab:top_pairs} the top-10 pairs of European airports which yield the highest classification score. Additionally, Fig. \ref{fig:rnd} in Appendix reports an analysis of the statistical significance of these results.

\Figure[t!][width=0.9\textwidth]{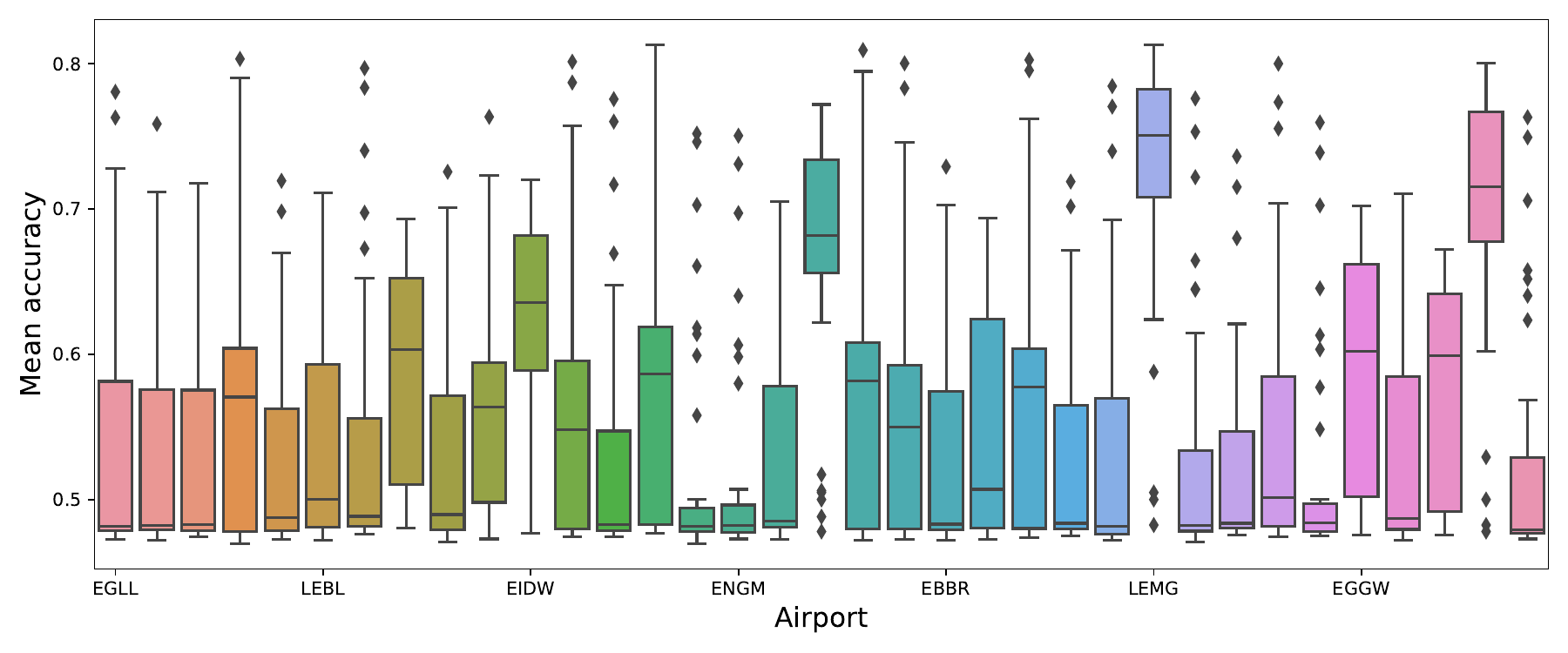}
{Overview of the classification of European delay propagation networks. Each box plot represents the distribution of the classification scores obtained between the networks of a given airport (X axis), and each other airport considered in this study. In each plot, the horizontal line represents the median, the box the interquartile range (IQR), and the whiskers the outliers outside the IQR. Airports are ordered in decreasing number of passengers, and as reported in Tab. \ref{tab:stats_airp}. \label{fig:boxplot}}

\begin{table*}
\caption{\label{tab:top_airps} Top-6 European airports with the highest average accuracy, i.e. whose delay propagation neighbourhood is easier to recognise. The right-most column reports the number of times each airport appears in the list of the $100$ pairs of airports with the highest classification score - see also Tab. \ref{tab:top_pairs}. }
\label{table}
\setlength{\tabcolsep}{3pt}
\center
\small
\begin{tabular}{|p{6.0cm}|p{1.3cm}|p{2.1cm}|p{1.9cm}|}
\hline
{\bf Name} & {\bf ICAO} & {\bf Mean accuracy} & {\bf Top-100 frequency}  \\ \hline
M\'alaga-Costa del Sol Airport & LEMG & $0.722$ & $28$ \\
Alicante-Elche Miguel Hern\'andez Airport & LEAL & $0.700$ & $27$ \\
London Stansted Airport & EGSS & $0.668$ & $22$ \\
Dublin Airport & EIDW & $ 0.624$ & $9$ \\
London Luton Airport & EGGW & $0.589$ & $9$ \\
London Gatwick Airport & EGKK & $0.585$ & $4$ \\ \hline
\end{tabular}
\end{table*}

\begin{table*}
\caption{\label{tab:top_pairs}Top-10 pairs of European airports yielding the highest mean accuracy, when classifying their respective neighbourhoods.}
\setlength{\tabcolsep}{3pt}
\center
\small
\begin{tabular}{|p{4.0cm}|p{4.0cm}|p{2.1cm}|}
\hline
{\bf Airport 1} & {\bf Airport 2} & {\bf Mean accuracy} \\ \hline
Palma de Mallorca Airport & M\'alaga-Costa del Sol Airport & 0.813 \\
Vienna International Airport & M\'alaga-Costa del Sol Airport & 0.809 \\
Frankfurt Airport & M\'alaga-Costa del Sol Airport & 0.803 \\
D\"usseldorf Airport & M\'alaga-Costa del Sol Airport & 0.803 \\
Z\"urich Airport & M\'alaga-Costa del Sol Airport & 0.801 \\ \hline
Palma de Mallorca Airport & Alicante-Elche Miguel Hern\'andez Airport & 0.800 \\
Stockholm Arlanda Airport & M\'alaga-Costa del Sol Airport & 0.800 \\
M\'alaga-Costa del Sol Airport & Hamburg Airport & 0.800 \\
Munich Airport & M\'alaga-Costa del Sol Airport & 0.797 \\
D\"usseldorf Airport & Alicante-Elche Miguel Hern\'andez Airport & 0.795 \\ \hline
\end{tabular}
\end{table*}

A simple visual inspection of the top-ranked airports in Tab. \ref{tab:top_airps} suggests that the identifiability of delay propagation neighbourhoods is related to low-cost carriers (LCCs). This is confirmed by the left panel of Fig. \ref{fig:corr_lcc}, reporting a scatter plot of the mean classification accuracy as a function of the proportion of flights operated by the four largest LCCs at each airport. A similar correlation can be observed when considering the proportion of flights operated by Ryanair, but not by the following three largest LCCs in Europe (see panels b-e). This is to be expected, as Ryanair was the first LCC in Europe in terms of number of operations in the considered time window, and hence dominates the results.

\Figure[t!][width=0.9\textwidth]{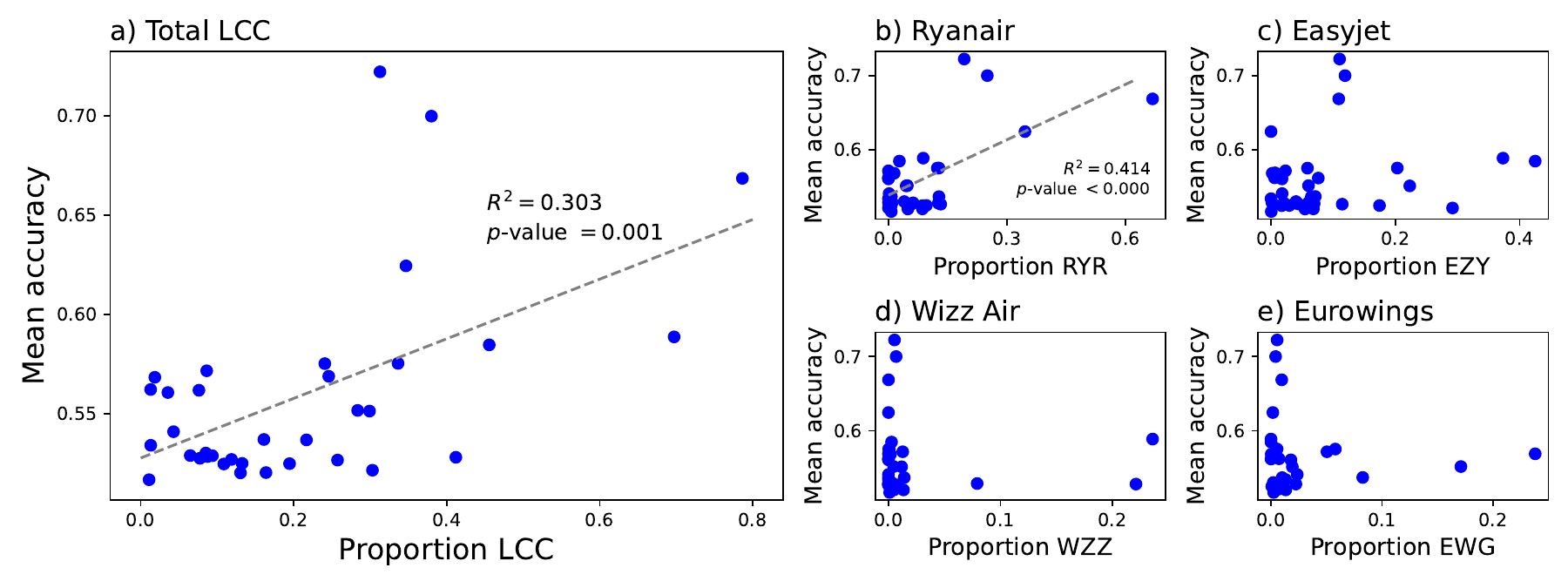}
{Role of low-cost carriers (LCCs) in the identifiability of delay propagation neighbourhoods. Panel a) reports a scatter plot of the mean accuracy as a function of the fraction of LCC operations at each airport; panels b) to e) the same information, disaggregated by the four largest LCCs operating in Europe. The dashed grey lines depict the best linear fit, with the corresponding $R^2$ and $p$-value being reported inside each panel. \label{fig:corr_lcc}}

We further analyse which other factors may affect the identifiability of the delay propagation neighbourhood of each airport. The top panels of Fig. \ref{fig:others} focus on the underlying physical connectivity networks, and specifically report the average classification accuracy of each pair of airports as a function of the fraction of common routes in their neighbourhoods (left panel), and of common destinations (right panel). As is to be expected, the more two airports share common destinations, the more difficult is to identify them - in the limit that, if two airports exactly share the same neighbours, their corresponding delay propagation neighbourhoods would be identical and it would be impossible to identify them. The bottom panels of the same figure evaluate the role of the identifiability of individual airports, i.e. how easy it is to recognise one airport given the evolution of the average delays of flights there operating - i.e. the identifiability of delays, as opposed to that of the delay propagation process. Identifiability values have been obtained from Ref. \cite{ivanoska2022assessing}, corresponding to classification tasks performed using Residual Networks (ResNet) Deep Learning models \cite{wang2017time, ismail2019deep}. The left bottom panel focuses on the local identifiability, i.e. how easy it is to identify the pair of airports under analysis; on the other hand, the right panel focuses on the neighbourhood (or meso-scale) identifiability, calculated as the average identifiability of all airports composing the neighbourhoods of the two airports under analysis. The identifiability of the central airport of each group is not a major driver (left panel), as is the average identifiability of constituting airports (right panel). Note that these results are to be expected, as here the focus is shifted towards the propagation of delays, as opposed to their local appearance - as measured in Ref. \cite{ivanoska2022assessing}.

\Figure[t!][width=0.9\textwidth]{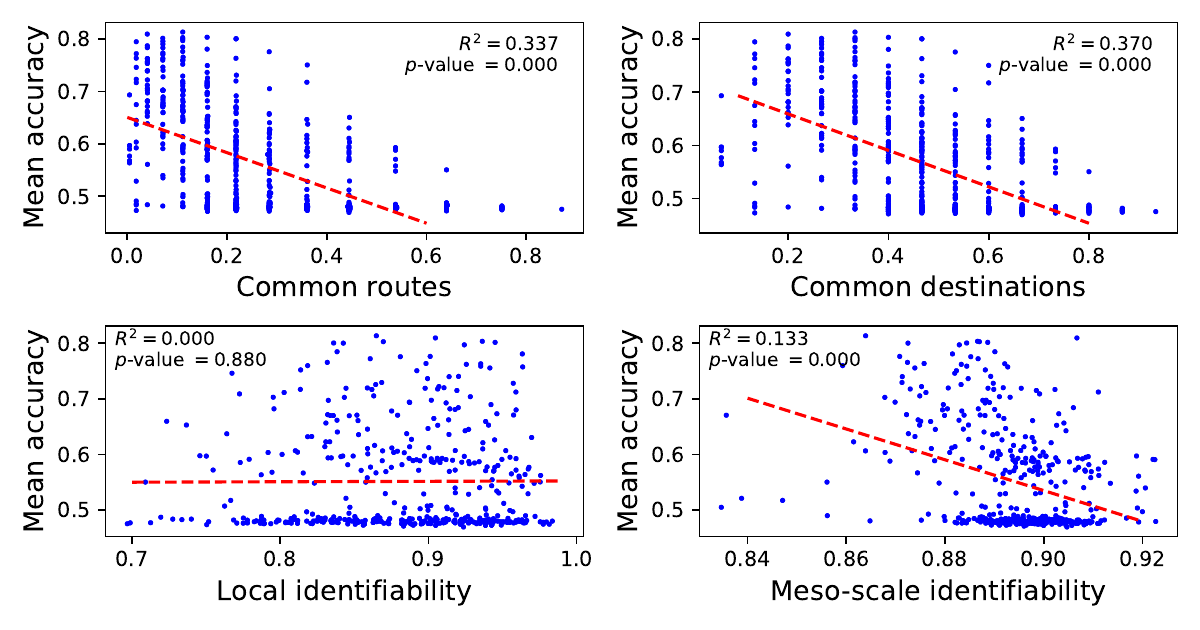}
{Additional factors affecting the identifiability of the delay propagation neighbourhoods of each pair of airports. From left to right, top to bottom, the four panels correspond to: the fraction of common routes; the fraction of common destinations; the local identifiability, i.e. how unique are the delay profiles of the two airports under study; and the average identifiability of all airports belonging to each neighbourhood. See main text for further details. The dashed red lines depict the best linear fit, with the corresponding $R^2$ and $p$-value being reported inside each panel. \label{fig:others}}

\subsection{Generalisability of results: functional and network metrics}
\label{sec:gen_func}

Some methodological choices made in the previous analyses may affect the obtained results. Specifically, we have opted for the use of a linear correlation in the reconstruction of the functional networks of delay propagation, due to the limited length of available time series; yet, other metrics could also provide complementary information. In order to check this aspect, the top panels of Fig. \ref{fig:other_metrics} report the main results obtained when using Granger Causality (top row) and Rank Correlation (middle row). From left to right, each column reports the results of optimising the threshold $\tau$, over the same three pairs of airports as in Fig. \ref{fig:hyperp}; the distribution of accuracies for each airport as box plots, as in Fig. \ref{fig:boxplot}; and a scatter plot of the accuracy obtained when classifying each pair of airports, as a function of the accuracy obtained with linear correlation. Some interesting conclusions can be drawn. On one hand, results obtained using the Granger Causality are not dissimilar from those of the linear correlation: the same relevant airports are detected, and only a handful of pairs of airport significantly improve the corresponding classification score. Therefore, even though the time series here available were very short, this causality metric leads to the same conclusions. On the other hand, the Rank Correlation fails at creating identifiable networks, for all airports and values of the threshold $\tau$. This suggests that the identifiability of delay propagation neighbourhoods is mostly driven by the linear part of the propagation; and that, when using a metric sensitive to non-linear relationships, the reconstructed networks are more random - see Fig. \ref{fig:RCRandom} in Appendix for a discussion.

\Figure[t!][width=0.9\textwidth]{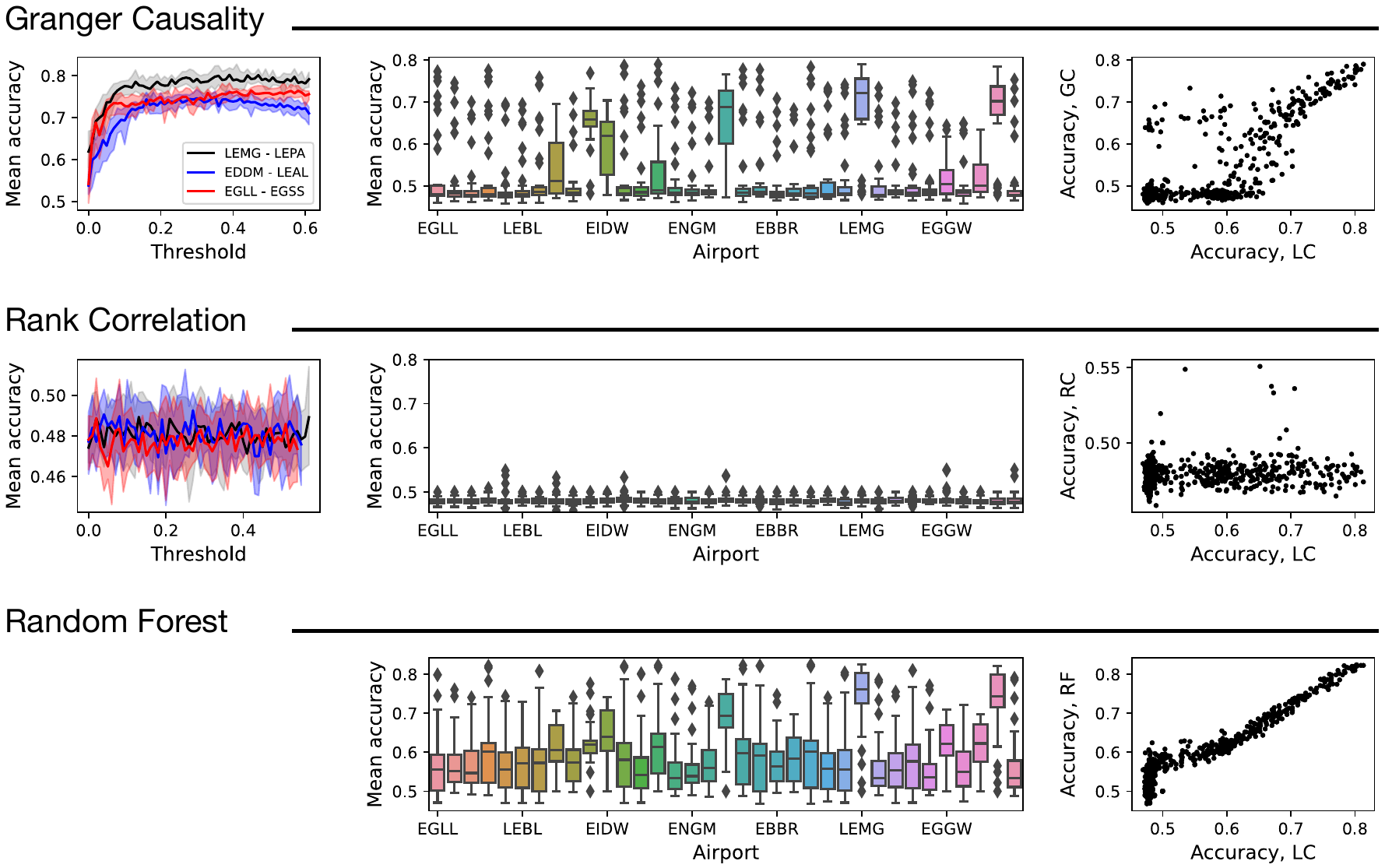}
{Alternative functional metrics and classification models. Top and middle rows report the results for delay propagation networks reconstructed using respectively Granger Causality (GC) and Rank Correlation (RC); the bottom row, for a classification performed using Random Forests on topological features extracted from the original networks. From left to right, each row reports the tuning of the threshold $\tau$; the distribution of the classification accuracy by airport; and a scatter plot, comparing these classification scores with the one obtained through linear correlation (LC) and Deep Learning. See main text for interpretations. \label{fig:other_metrics}}

This work further relies on Deep Learning models to estimate the identifiability of the reconstructed networks; a pertinent question is whether these are needed, and whether they introduce some distortion in the results. As a reference, we have tested the performance of a classical Machine Learning algorithm, i.e. Random Forests (RFs). These are based on the concept of Decision Trees, i.e. comprehensive tree structures that classify records by sorting them based on attribute values \cite{priyam2013comparative}; and further expand on this idea by combining a large number of tree predictors, such that each tree depends on the values of a random vector sampled independently and with the same distribution for all trees in the forest \cite{breiman2001random}. The analysis entails, firstly, extracting a set of classical topological metrics from each network, i.e. measures describing some specific aspects of its structure - see Sec. \ref{sec:top_metrics} for the full list, and Fig. \ref{fig:metrics} in Appendix for an example involving two airports. Secondly, the corresponding classification score has been calculated using RFs, trained and evaluated over those topological features. The results (bottom row of Fig. \ref{fig:other_metrics}) indicate that very similar results can be obtained; using Deep Learning models does not provide any clear advantage, but also introduces no bias.

\subsection{Generalisability of results: the case of US}
\label{sec:gen_US}

In order to understand if the aforementioned results are specific to the EU or more general, we have performed a similar analysis on data representing the dynamics at US airports. In order to simplify the comparison, we have maintained the same methodological choices, i.e. the use of linear correlations and $n_{airps} = 15$. We have nevertheless optimised the threshold $\tau$, obtaining that values higher than what previously used help in the classification, with a maximum for $\tau = 0.45$ - see left panel of Fig. \ref{fig:US}.

The right panel of Fig. \ref{fig:US} further reports the probability distributions of the identifiability of each airport as box plots, i.e. akin to Fig. \ref{fig:boxplot} - see also the right panel of Fig. \ref{fig:all_pairs} in Appendix for individual values.
Classification scores are in general much lower than in the EU case. The reason for this cannot be identified at this stage, although the larger size of the data can be excluded as a cause, see Fig. \ref{fig:downs} in Appendix.
Among all neighbourhoods, only three airports stand out: San Jos\'e International Airport (SJC), Fort Lauderdale-Hollywood International Airport (FLL) and Chicago O'Hare International Airport (ORD). It is worth noting that these three airports are the headquarters or handle a large proportion of flights of several low-cost and ultra low-cost airlines, respectively: Southwest Airlines (SJC); Allegiant Air, JetBlue and Spirit Airlines (FLL); and Spirit Airlines (ORD). The importance of such type of airlines in the identifiability of delay propagation neighbourhoods seems therefore to be confirmed also in the case of US.

\Figure[t!][width=0.9\textwidth]{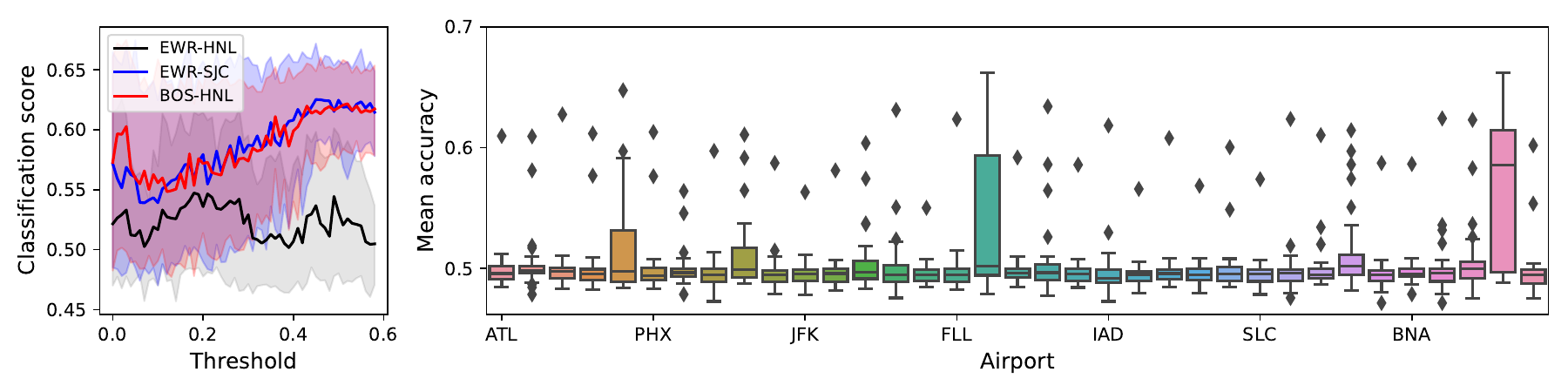}
{Identifiability of US delay propagation neighbourhoods. The left panel reports the evolution of the average accuracy of the classification as a function of the threshold $\tau$, for three pairs of airports. Solid lines depict the average classification score, transparent bands the corresponding $10-90$ percentiles. The right panel represents the distribution of the classification scores obtained between the networks of a given airport (X axis), and all other US airports considered in this study. \label{fig:US}}

\section{Discussion and conclusions}
\label{sec:concl}

In this contribution we have proposed an analysis of the identifiability (or uniqueness) of the delay propagation patterns developing in the neighbourhood of a given airport. This represents an evolution of a previous research work that assessed the identifiability of delays profiles at individual airports \cite{ivanoska2022assessing}, thus moving the focus from the dynamics of a single element to the coordinated dynamics of many of them. From a methodological viewpoint, this has been achieved by leveraging Deep Learning models, and specifically Graph Isomorphism Neural Networks (GIN) \cite{Xu_Hu_Leskovec_Jegelka_2019}, on functional networks of delay propagation obtained through linear correlation; yet, results are robust against other methodological choices - see Fig. \ref{fig:other_metrics}. To the best of our knowledge, this is the first instance of a study focusing on how delays propagate in the neighbourhood of an airport; and one of the few examples of the use of DL models beyond simple classifications.

Obtained results are highly heterogeneous; while most airports are not identifiable, a few of them yield very high classification scores, indicating that the propagation of delays in their neighbourhoods is highly unique - see Fig. \ref{fig:boxplot}. A simple visual inspection of these airports (Tabs. \ref{tab:top_airps} and \ref{tab:top_pairs}) suggests that such uniqueness is related to the predominance of Low Cost Carriers (LCCs), an intuition that has numerically been checked in Fig. \ref{fig:corr_lcc}. Additionally, and as it may be expected, pairs of airports are easier to be identified when their physical connectivity structure (i.e. the set of destinations they have access to) is substantially different, see Fig. \ref{fig:others}. Finally, the importance of LCCs is not a peculiarity of the European air transport system, but also applies to US, see Fig. \ref{fig:US}.

LCCs are known to mostly (but not always \cite{wu2020network}) operate routes based on a point-to-point structure \cite{dobruszkes2006analysis, flores2009airline, dobruszkes2013geography, lordan2014study}, as opposed to the hub-and-spoke networks of traditional airlines.
Such difference in the way flights are scheduled has a direct impact on delays. On one hand, LCCs contribute to a reduction of delays at individual airports \cite{rupp2008low, bubalo2015low}, i.e. low fares do not necessarily equate low quality of service. On the other hand, they are more prone to propagate delays, due both to their business model of quick aircraft turnarounds \cite{brueckner2022airline}, and to the lack of hubs, where carriers have more available resources and can internalise delays \cite{bubalo2021flight}. Results here presented suggest that these differences have a clear impact in the way delays are propagated: airports with a high percentage of flights operated by LCCs are embedded in delay propagation sub-networks that are different and differentiable from those of airports served by traditional airlines.

It is important to note that an identifiable propagation sub-network is not the same as a different, or isolated, sub-network of connections. As illustrated in the top right panel of Fig. \ref{fig:others}, airports that share few destinations are naturally easier to be differentiated, but only on average. It is not difficult to find pairs of airports with almost no common destinations, and still not being identifiable, suggesting that these destinations are themselves embedded in common delay propagation patterns. Similarly, pairs of airports with a high share of LCC flights are themselves identifiable, see for instance the case of Palma de Mallorca Airport and M\'alaga-Costa del Sol Airport; LCCs thus induce heterogeneous delay propagation patterns that are different depending on the considered airport. Finally, the identifiability of the neighbourhood networks does not strongly depend on the central airport itself, but is mostly defined by the propagation patterns between neighbours (see Fig. \ref{fig:no_central} in Appendix); what is important is not therefore how delays are generated in the central airport, but rather how they then propagate between neighbours.

The concept here proposed can help improving our understanding of delays and their propagation. If delays were the unavoidable consequence of random events in the system, as e.g. of a technical failure reducing the capacity of an airport and consequently creating delays, they should appear as random events themselves. In a way akin to the Efficient Market Hypothesis in finance \cite{malkiel2003efficient}, the fact that delays can be forecast (i.e. that they are not completely random) implies an inefficiency in the system. To illustrate this point with an trivial example, if flights are always delayed in a specific day and hour, one may wonder why nothing is done to prevent that. Additionally, if delays are identifiable, this implies that they also are predictable - as identifiability requires the presence of unique and constant patterns in the data. The question thus becomes, at what level are delays predictable and/or identifiable? Previous literature has shown that delays are predictable at the scale of individual flights \cite{rebollo2014characterization, yu2019flight, cai2021deep}; and that they are further predictable and identifiable at airport level \cite{ivanoska2022assessing}. We here introduce an additional element: their propagation is also predictable and identifiable at a supra-airport level.

Major changes in the number of LCC flights operating at an airport seem to result in different delay propagation patterns among neighbours. This constitute an indirect effect: the identifiability does not come from those LCC flights directly, but rather from how the delays by them generated are further propagated. From an operational perspective, it highlights the networked nature of the problem: the delays an airport has to manage do not only depend on its operations, but also on the operations of its neighbours. Hence, consequences of any change in traffic patterns must be evaluated on a meso- and macro-scale level. Still, it has also to be noted that results here presented do not demonstrate causality, but rather correlations; and that further research is needed to better understand the mechanisms behind identifiability - including airport/airspace configurations and air traffic management procedures.


\appendices

\begin{table*}
\caption{\label{tab:stats_airp}Information on the $35$ European airports considered in this study, including their 4-letters ICAO code and the number of flights landed in the considered time period.}
\setlength{\tabcolsep}{3pt}
\center
\small
\begin{tabular}{|p{0.9cm}|p{7.0cm}|p{1.2cm}|p{1.6cm}|}
\hline
{\bf Rank} & {\bf Name} & {\bf ICAO} & {\bf \# landings}  \\ \hline
$1$ & London Heathrow & EGLL & $392,911$ \\
$2$ & Paris Charles de Gaulle Airport & LFPG & $400,465$ \\
$3$ & Amsterdam Airport Schiphol & EHAM & $406,303$ \\
$4$ & Frankfurt Airport & EDDF & $400,181$ \\
$5$ & Adolfo Su\'arez Madrid-Barajas Airport & LEMD & $325,812$ \\ \hline
$6$ & Josep Tarradellas Barcelona-El Prat Airport & LEBL & $264,199$ \\
$7$ & Munich Airport & EDDM & $329,424$ \\
$8$ & London Gatwick Airport & EGKK & $234,931$ \\
$9$ & Rome-Fiumicino International Airport & LIRF & $255,442$ \\
$10$ & Paris Orly Airport & LFPO & $192,496$ \\ \hline
$11$ & Dublin Airport & EIDW & $181,877$ \\
$12$ & Z\"urich Airport & LSZH & $210,029$ \\
$13$ & Copenhagen Kastrup Airport & EKCH & $217,833$ \\
$14$ & Palma de Mallorca Airport & LEPA & $163,508$ \\
$15$ & Humberto Delgado Airport & LPPT & $163,891$ \\ \hline
$16$ & Oslo Airport & ENGM & $204,646$ \\
$17$ & Manchester Airport & EGCC & $161,061$ \\
$18$ & London Stansted Airport & EGSS & $148,841$ \\
$19$ & Vienna International Airport & LOWW & $206,649$ \\
$20$ & Stockholm Arlanda Airport & ESSA & $198,036$ \\ \hline
$21$ & Brussels Airport & EBBR & $186,183$ \\
$22$ & Milan Malpensa Airport & LIMC & $156,061$ \\
$23$ & D\"usseldorf Airport & EDDL & $177,683$ \\
$24$ & Athens Intl Eleftherios Venizelos & LGAV & $158,173$ \\
$25$ & Berlin Tegel ``Otto Lilienthal'' Airport & EDDT & $151,402$ \\ \hline
$26$ & M\'alaga-Costa del Sol Airport & LEMG & $101,539$ \\
$27$ & Warsaw Chopin Airport & EPWA & $136,973$ \\
$28$ & Geneva Airport & LSGG & $143,293$ \\
$29$ & Hamburg Airport & EDDH & $121,395$ \\
$30$ & Václav Havel Airport Prague & LKPR & $116,156$ \\ \hline
$32$ & Budapest Ferenc Liszt International Airport & LHBP & $85,870$ \\
$33$ & Edinburgh Airport & EGPH & $102,110$ \\
$34$ & Alicante-Elche Miguel Hern\'andez Airport & LEAL & $73,539$ \\
$35$ & Nice C\^ote d'Azur Airport & LFMN & $107,589$ \\ \hline
\end{tabular}
\end{table*}

\begin{table*}
\caption{\label{tab:stats_airp_US}Information on the $35$ US airports considered in this study, including their 3-letters IATA code and the number of flights landed in the considered time period.}
\setlength{\tabcolsep}{3pt}
\center
\small
\begin{tabular}{|p{0.9cm}|p{8.5cm}|p{1.2cm}|p{1.6cm}|}
\hline
{\bf Rank} & {\bf Name} & {\bf ICAO} & {\bf \# landings}  \\ \hline
$1$ & Hartsfield-Jackson Atlanta International Airport & ATL & $209,835$ \\
$2$ & Denver International Airport & DEN & $145,416$ \\
$3$ & Dallas Fort Worth International Airport & DFW & $147,873$ \\
$4$ & Los Angeles International Airport & LAX & $181,800$ \\
$5$ & Chicago O'Hare International Airport & ORD & $174,802$ \\ \hline
$6$ & Phoenix Sky Harbor International Airport & PHX & $125,039$ \\
$7$ & Minneapolis-Saint Paul International Airport & MSP & $82,023$ \\
$8$ & Charlotte Douglas International Airport & CLT & $93,509$ \\
$9$ & Seattle-Tacoma International Airport & SEA & $93,741$ \\
$10$ & San Francisco International Airport & SFO & $129,724$ \\ \hline
$11$ & John F. Kennedy International Airport & JFK & $83,528$ \\
$12$ & George Bush Intercontinental Airport & IAH & $95,204$ \\
$13$ & Orlando International Airport & MCO & $98,950$ \\
$14$ & Newark Liberty International Airport & EWR & $86,113$ \\
$15$ & Harry Reid International Airport & LAS & $121,662$ \\ \hline
$16$ & Fort Lauderdale-Hollywood International Airport & FLL & $69,246$ \\
$17$ & General Edward Lawrence Logan International Airport & BOS & $105,598$ \\
$18$ & Detroit Metropolitan Wayne County Airport & DTW & $81,457$ \\
$19$ & Miami International Airport & MIA & $68,185$ \\
$20$ & LaGuardia Airport & LGA & $89,140$ \\ \hline
$21$ & Washington Dulles International Airport & IAD & $35,400$ \\
$22$ & Baltimore/Washington International Thurgood Marshall Airport & BWI & $65,822$ \\
$23$ & Philadelphia International Airport & PHL & $68,386$ \\
$24$ & San Diego International Airport & SAN & $71,672$ \\
$25$ & Chicago Midway International Airport & MDW & $70,031$ \\ \hline
$26$ & Salt Lake City International Airport & SLC & $69,190$ \\
$27$ & Ronald Reagan Washington National Airport & DCA & $73,611$ \\
$28$ & Tampa International Airport & TPA & $60,144$ \\
$29$ & Portland International Airport & PDX & $45,353$ \\
$30$ & St. Louis Lambert International Airport & STL & $47,467$ \\ \hline
$31$ & Nashville International Airport & BNA & $48,158$ \\
$32$ & Austin-Bergstrom International Airport & AUS & $43,136$ \\
$33$ & Daniel K. Inouye International Airport & HNL & $18,273$ \\
$34$ & San Jos\'e International Airport & SJC & $36,353$ \\
$35$ & Kansas City International Airport & MCI & $40,626$ \\ \hline
\end{tabular}
\end{table*}

\Figure[t!][width=0.9\textwidth]{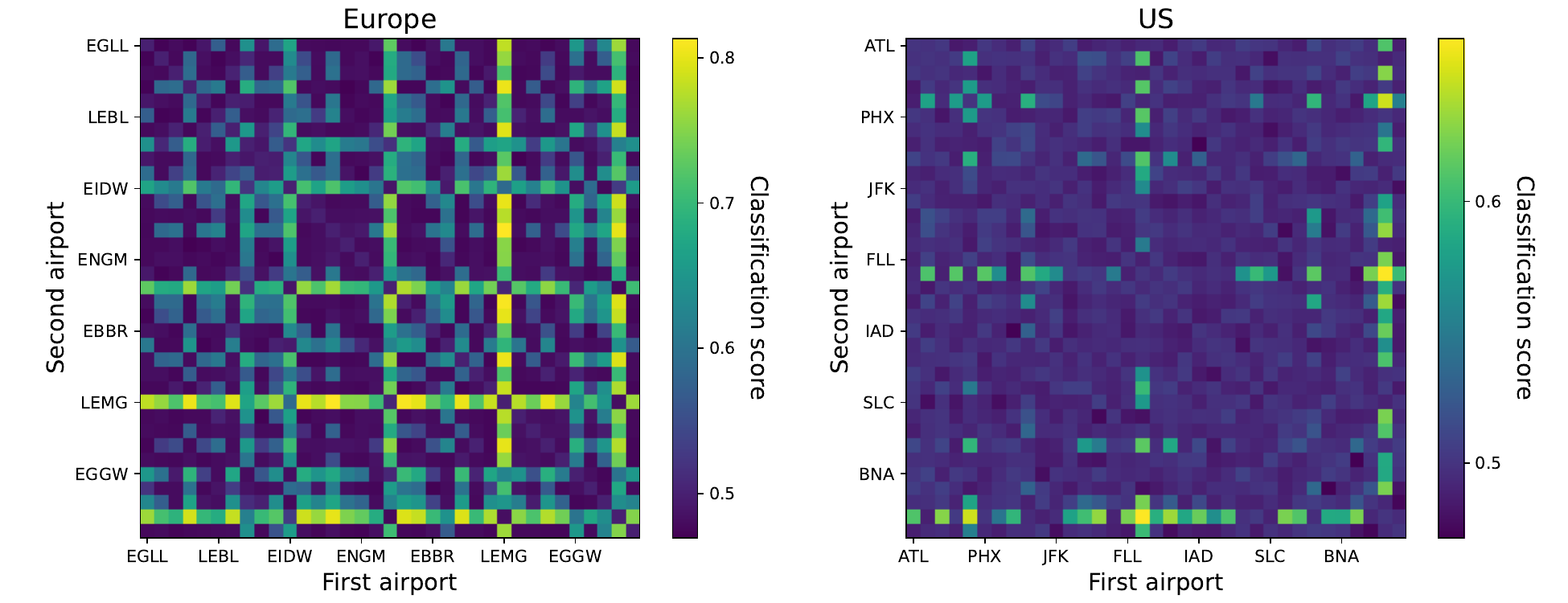}
{Classification scores between all delay propagation sub-networks centred in all pairs of airports in EU (left) and US (right), obtained using GIN models. Airports are sorted (from left to right, and top to bottom) in increasing order of passengers - see the full lists in Tabs. \ref{tab:stats_airp} and \ref{tab:stats_airp_US}. \label{fig:all_pairs}}

\Figure[t!][width=0.9\textwidth]{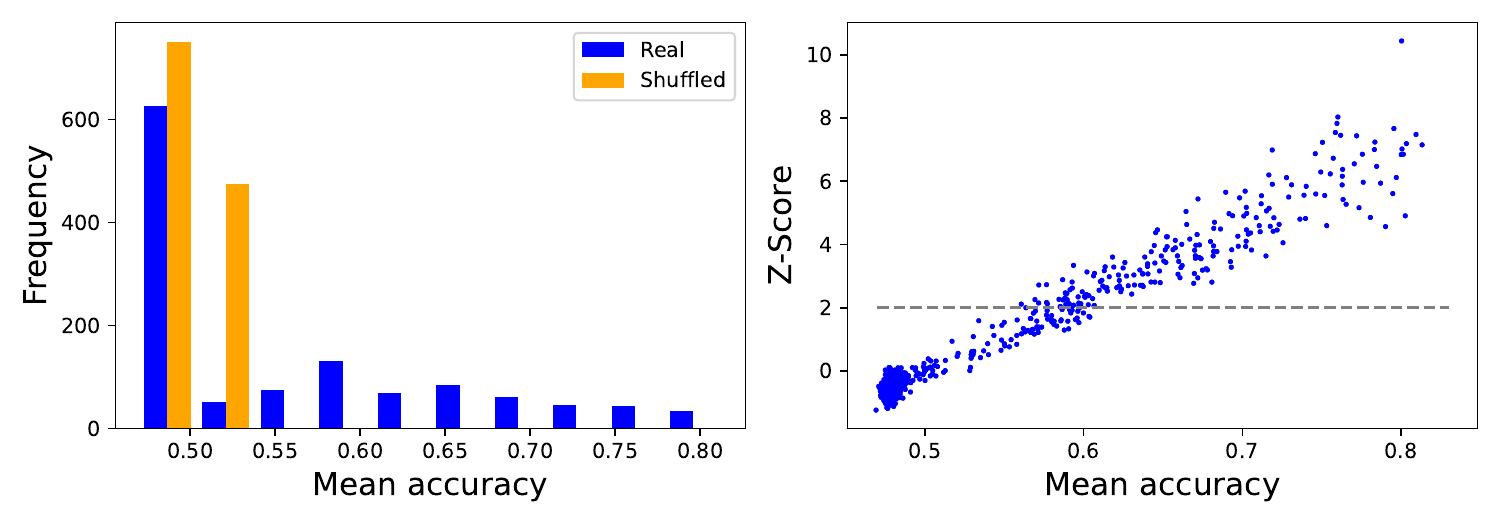}
{Analysis of the statistical significance. The left panel reports histograms of the distribution of the mean accuracy of the classification score of each pair of airports, using the real data (blue bars) and a version of the same with labels being randomly shuffled (orange bars). As is to be expected, in the latter case the accuracy stays around $0.5$, thus indicating a random (or non-informative) classification, and thus the absence of overfitting. The right panel further reports the evolution of the Z-Score of the accuracy of each pair of airports, calculated against the score obtained in the shuffled case. In other words, this metric reports how much the obtained score deviates from what obtained in data without any structure. All pairs yielding an accuracy greater than $0.6$ also have a Z-Score greater than two, thus indicating that these results are statistically significant. \label{fig:rnd}}

\Figure[t!][width=0.9\textwidth]{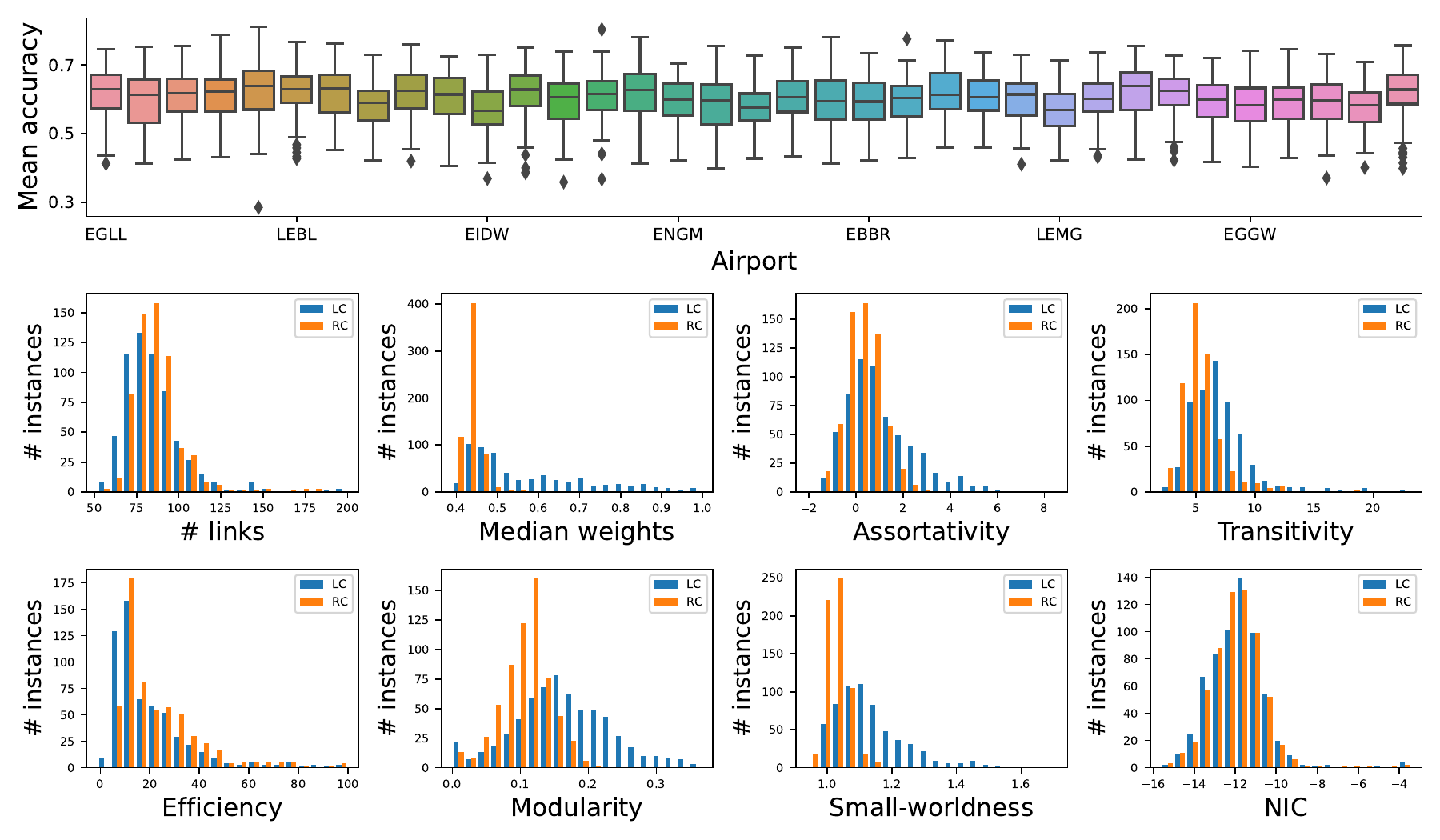}
{Analysis of networks reconstructed using Rank Correlation (RC). The top panel reports the distribution of the classification score obtained when delay propagation neighbourhoods of each airports have been classified against randomised version of the same. The bottom panels depict histograms of the values of eight standard topological metrics, calculated on the delay propagation neighbourhoods for Palma de Mallorca Airport (LEPA) obtained using Linear Correlation (LC, blue bars) and Rank Correlation (RC, orange bars). A description of the considered topological metrics can be found in Sec. \ref{sec:top_metrics}. It can be appreciated that topological metrics in the RC case are closer to what expected in random networks, with smaller assortativity, transitivity, modularity and small-worldness. This is further confirmed by the classification results in the top panel: the obtained networks are very similar to random ones, hindering the classification task. \label{fig:RCRandom}}

\Figure[t!][width=0.9\textwidth]{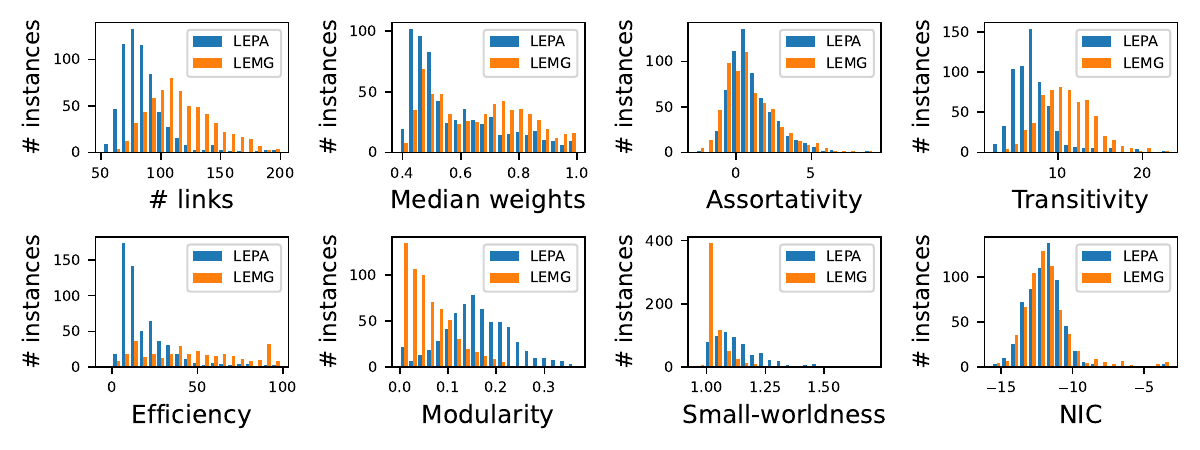}
{Topological metrics extracted from delay propagation sub-networks reconstructed using Linear Correlation. Each panel reports the histogram of a different metric, for all networks extracted from Palma de Mallorca Airport (LEPA, blue bars) and M\'alaga-Costa del Sol Airport (LEMG, orange bars), i.e. the pair of airports with the highest identifiability. See Sec. \ref{sec:top_metrics} for a description of the topological metrics. Clear differences can be observed, especially for the number of links, transitivity, and modularity. \label{fig:metrics}}

\Figure[t!][width=0.9\textwidth]{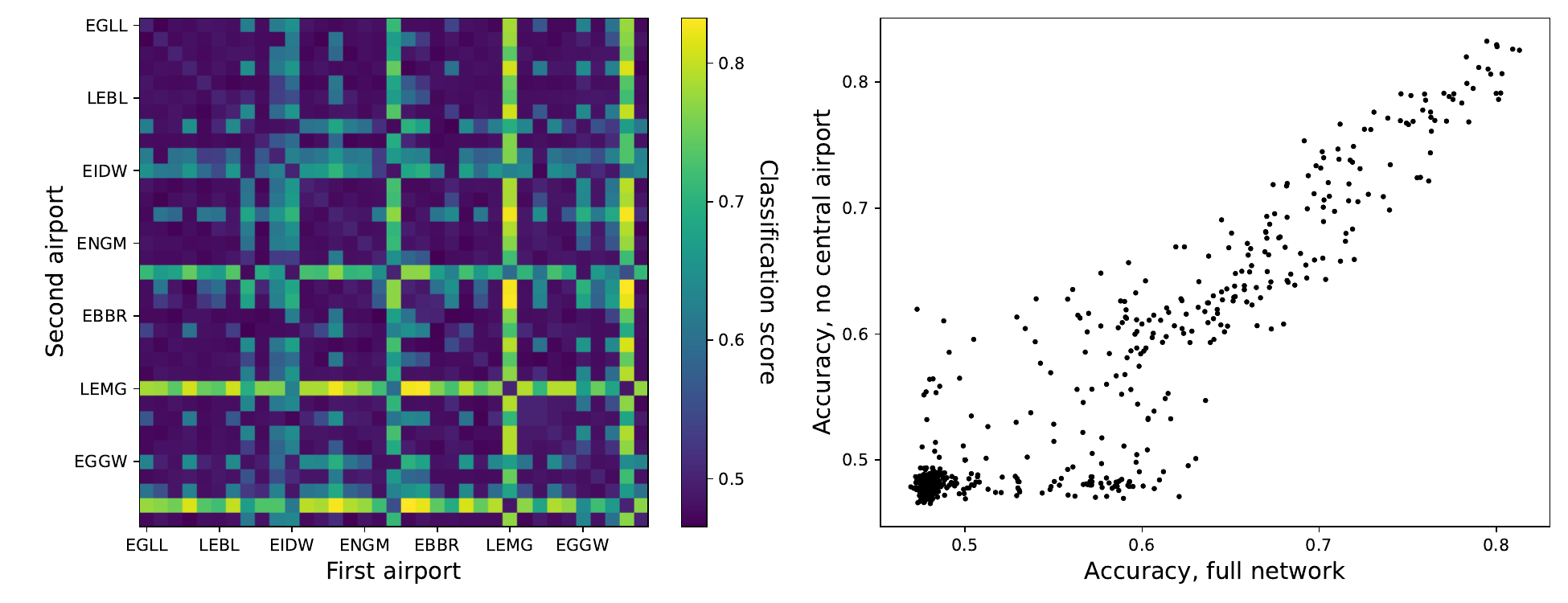}
{Impact of the central airport in the classification score. The left panel reports the classification score obtained when the central airport of each neighbourhood is deleted, including all its connections. In other words, we here analyse whether the central airport has a major role in the identifiability of the corresponding sub-network; or whether, on the contrary, this is defined by the connectivity patterns among the neighbours. It can be appreciated that the structure observed here is qualitatively the same as in the left panel of Fig. \ref{fig:all_pairs}, thus supporting the latter explanation. The scatter plot in the right panel, comparing the score with (X axis) and without (Y axis) central node, and specifically the fact that points are clustered along the main diagonal, further confirms the previous intuition. \label{fig:no_central}}

\Figure[t!][width=0.9\textwidth]{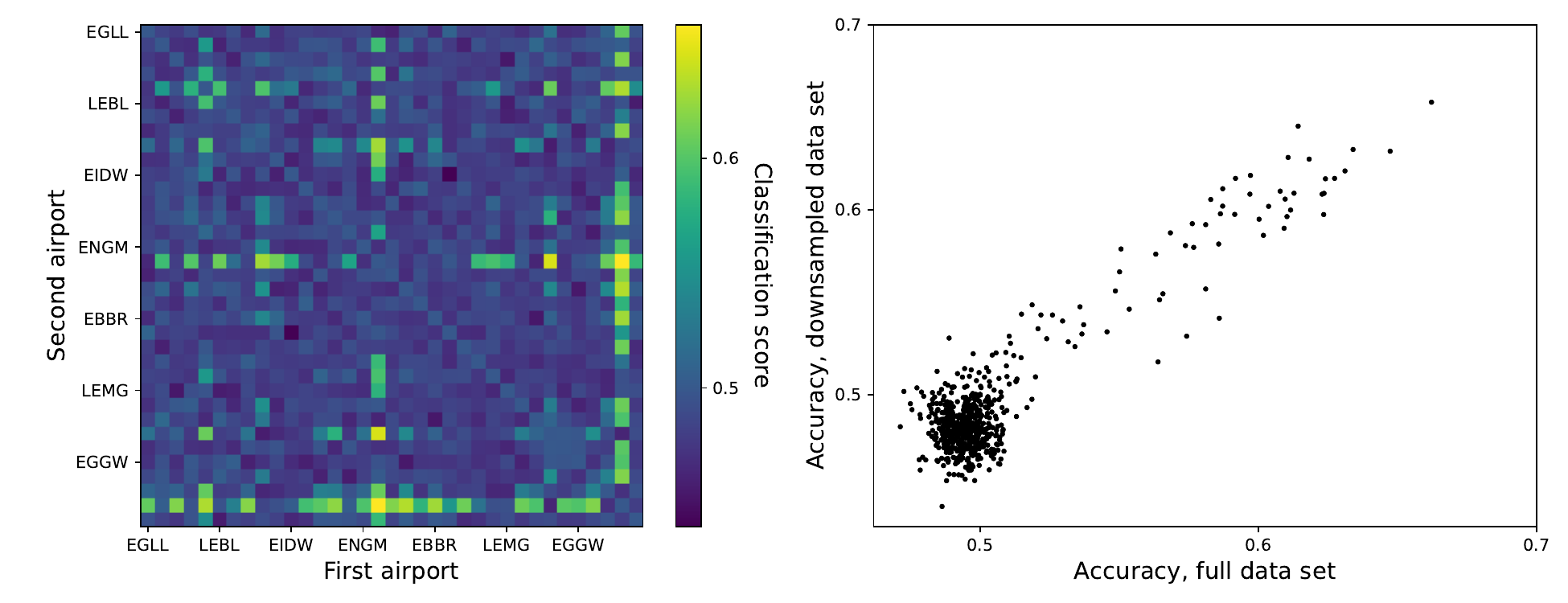}
{Impact of the size of the data set. The left panel reports the classification score between all pairs of US airports when the model is trained and tested on a subset of the original data, comprising a number of randomly selected days equivalent to what available in the EU case. The scatter plot in the right panel, comparing the accuracy score obtained with the complete and reduced data sets, excludes the size of the data as a potential reason for the lower identifiability observed in the US case. \label{fig:downs}}

\clearpage

\bibliographystyle{IEEEtran}

\bibliography{DelayNetIdent.bib}

\begin{IEEEbiography}[{\includegraphics[width=1in,height=1.25in,clip,keepaspectratio]{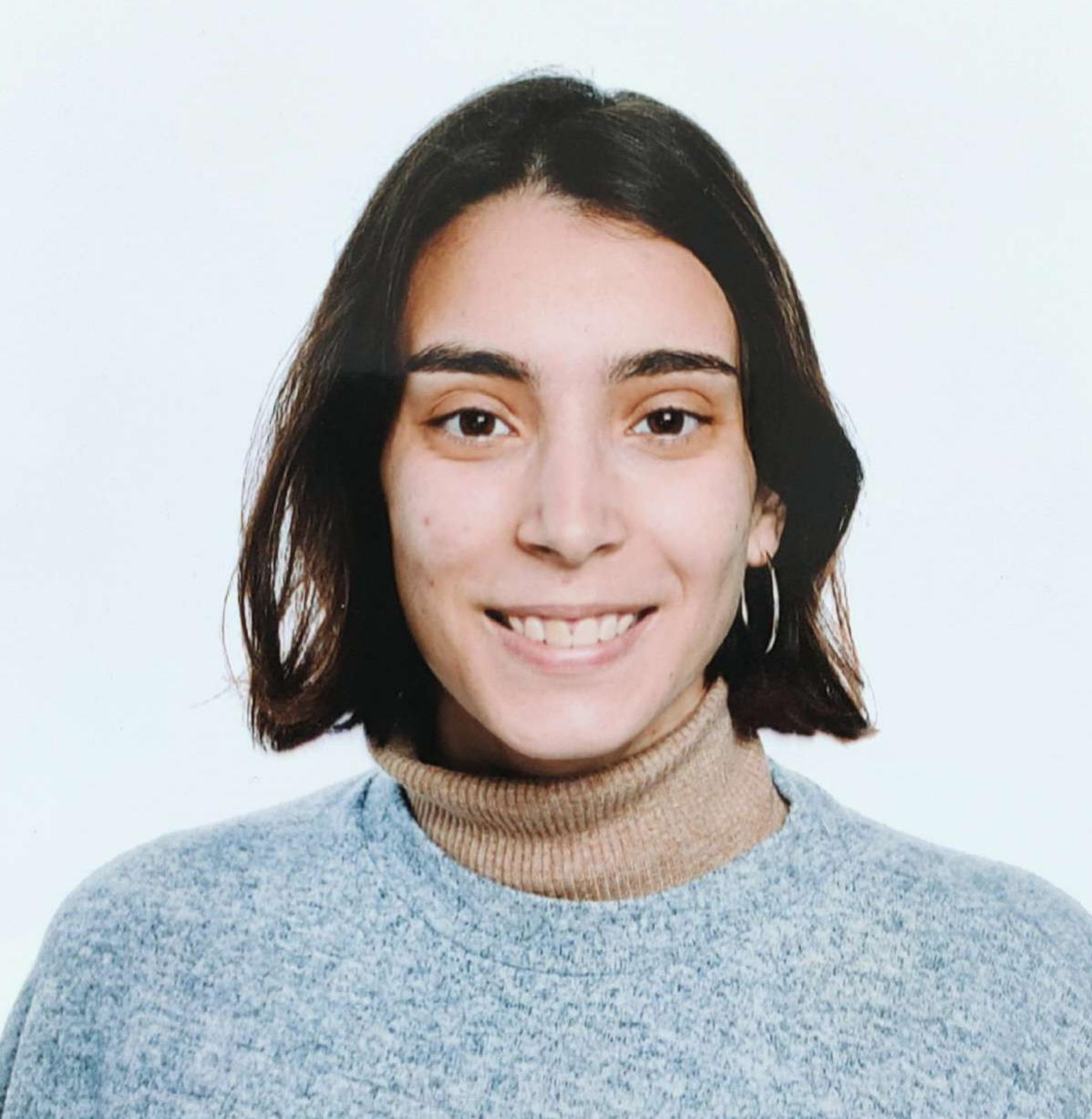}}] {Sofia Gil-Rodrigo} was born in València, Spain, in 2001. She graduated in Physics from the University of València in 2023. She is currently finishing the Master's degree in Physics of Complex Systems while collaborating with a research scholarship in the Institute for Cross-Disciplinary Physics and Complex Systems, Palma de Mallorca, Spain.

She focuses on the study of complex networks from different complex systems, including delay propagation in air transport, using Deep Learning methods.
\end{IEEEbiography}

\begin{IEEEbiography}[{\includegraphics[width=1in,height=1.25in,clip,keepaspectratio]{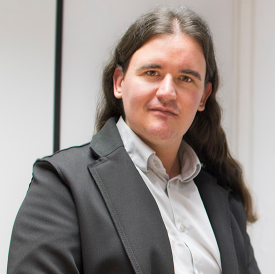}}] {Massimiliano Zanin} was born in Verona, Italy, in 1982. He received the Ph.D. on computer engineering from the Universidade Nova de Lisboa, Portugal, in 2014.

He is currently a researcher at the Institute for Cross-Disciplinary Physics and Complex Systems in Palma de Mallorca, Spain. His main topics of interest are Complex Networks and Data Science, both from a theoretical perspective and through their application to several real-world problems. Throughout his career he published more than 130 papers in international journals, and more than 50 contributions in conferences, reaching an h-index of 37. He has participated in several European competitive research projects, and he is currently the PI of the ERC StG ARCTIC, on the analysis and modelling of delay propagation in air transport.
\end{IEEEbiography}

\EOD

\end{document}